\documentstyle [aps,prl,psfig,twocolumn,epsf]{revtex}

\def\p{{\bf p}}
\def\k{{\bf k}}

\def\r{{\bf r}}
\def\rp{{\bf r^\prime}}

\begin{document}

\title{A natural orbital method for the electron
momentum distribution in matter}

\author{Bernardo Barbiellini}

\address{
    Physics Department,
    Northeastern University, 
    Boston, Massachusetts 02115 }

\date{\today}
\maketitle

\begin{abstract}
A variational method for many electron
system is applied to momentum distribution calculations.
The method uses a generating two-electron geminal and  
the amplitudes of the occupancies of one particle
natural orbitals as variational parameters. It introduces 
correlation effects beyond the free fermion
nodal structure.
\end{abstract}

\pacs{
PACS numbers: 31.25.-v, 71.10.-w
%31.25.-v Electron correlation calculations for atoms and molecules
%71.10.-w Theories and models of many electron systems
}

\section*{}
The characteristics of condensed matter systems are due to the motion 
and correlation of the electrons \cite{fulde}. 
The electron motion can be observed
by Compton scattering with photons or by positron annihilation.
Recent experiments \cite{sakurai,alfred} indicate that the momentum
density even in simple metals cannot be well represented by a single 
Slater determinant state.
Instead, the momentum density has to be constructed from a
correlated state with average occupancies $n_i$ of single particle
states in between 0 and 1 \cite{kubo}.
In other words, at the independent particle level there are only
$N$ occupancies different from zero, and these are
equal to one. When correlation is introduced, we have an infinite
number of occupancies different from zero, even though
most of them will presumably be very close to zero.
The sum rule
\begin{equation}
N=2~\sum_i n_i
\end{equation}
remains always true (the factor 2 is due to the spin). 
The purpose of the present work is to give a simple and efficient
calculation method to estimate the occupancies $n_i$.
To simplify the problem, we will consider approximations
which neglect the spin. Therefore the present discussion applies
to non-magnetic systems.

If the many body state is given by the wave function $\Psi$,
the first-order density matrix $\rho$ is defined by
\begin{equation}
\rho(\r,\rp)=N\int d\xi ~\Psi^*(\r,\xi)\Psi(\rp,\xi)~.
\end{equation}
The eigenfunctions $\psi_i$ of $\rho$,
introduced by L\"owdin as {\em natural orbitals} \cite{lowdin}
are the most suitable set of one-particle functions to use
in this discussion 
\begin{equation}
\rho=2\sum_i n_i~|~\psi_i><~\psi_i~|~.
\end{equation}
The natural orbitals form an orthonormal basis set.
Another set of orbitals that are naturally associated 
with many-body functions are the {\em generalized
overlap amplitudes} \cite{goscinki}.
These orbitals are, however, linearly dependent and 
a canonical orthonormalization of them yields the
natural orbitals.

The range of the the first-order density matrix $\rho(\r,\rp)$ in real space 
is a fundamental property of quantum mechanical systems 
since it determines
the degree of locality of the bonding properties \cite{kohn}. 

The two-particle reduced density matrix $\sigma$ 
contains all the information to discuss two-particle interactions $V_2$
\cite{march,coleman} and the total energy can be expressed as
\begin{equation}
E[\sigma]=(N/2)~{\rm Tr}(K \sigma)~,
\end{equation}
\begin{equation}
K(\r_1,\r_2)=H_1(\r_1)+H_1(\r_2)+(N-1)V_2(\r_1,\r_2)~,
\end{equation}
where $H_1$ is the one-body part of the hamiltonian.
If $\Psi$ is given by single Slater determinant,
as in the Hartree-Fock approximation or in the density 
functional theory \cite{kohn}, then
the energy is even determined by a one-particle
density matrix $\rho$, such as $\rho=\rho^2$ (idempotency).
Recently, Goedecker and Umrigar (GU) \cite{stefan} proposed to 
relax the $\rho$ idempotency and to use a 
a natural orbital functional for $\sigma$. 
The GU functional gives still
a particular importance to the individual electron picture.

In the present work,
an alternative method is explored.
One considers the ansatz proposed by Blatt 
\cite{blatt,bouchard,agp},
\begin{equation}
\Psi=\mbox{const}~[
\sum_i~g_i~a^+(\psi_i^*)a^+(\psi_i)]^{N/2}~|~0~>.
\end{equation}
where $a^+(\psi_i)$ are creation
operators of an electron in the state $\psi_i$.
In coordinates space, $\Psi$ 
is an Antisymmetrized Geminal Product (AGP) 
\begin{equation}
\Psi=\mbox{const}~\mbox{Det}|\phi(\r_i-\r_j)|.
\end{equation}
The {\em generating geminal} $\phi$ has a diagonal expansion in the
natural orbitals
\begin{equation}
\phi(\r_1,\r_2)=\sqrt{\frac{2}{N}}
\sum_i~g_i~\psi_i^*(\r_1) \psi_i(\r_2)~.
\end{equation}

In practice, the total energy becomes a functional $E[g_i,\psi_i]$.
Thus, $g_i$ and $\psi_i$ are determined by minimizing this functional. 
Such calculations have been done for some
molecules \cite{kurtz}.
The Stochastic Gradient Approximation (SGA) optimization
\cite{sga} is particularly appropriate for the present 
problem  since
the variational parameters can be determined
avoiding the explicit
determination of the total energy. 

The AGP is the $N$ particle component of the  
the BCS state \cite{schrieffer}.
In the limit of $N$ large,
the AGP and BCS states become identical.
If one set
$g_i=v_i/u_i$
with $|u_i|^2+|v_i|^2=1$,
then $n_i=|u_i|^2$. Therefore the electron momentum distribution 
$n(\p)$ is given by the simple formula \cite{barnes_bcs}
\begin{equation}
n(\p)=2\sum_i |v_i|^2~|<\p~|~\psi_i>|^2~.
\end{equation}
The expectation value in the AGP of two-particle operators 
can be found in ref. \cite{agp}. 

For a 2 electron system, the present scheme is equivalent
to a configuration interaction calculation and 
the two-particle reduced density matrix
$\sigma$ is given by a pure state $|\phi>$
\begin{equation}
\sigma=|~\phi><~\phi~|,
\end{equation}
thus $\sigma=\sigma^2$.

The hydrogen molecule is a good example to illustrate the method.
The bonding and antibonding orbitals are 
\begin{equation}
\psi_0 (\r) = {1 \over [2(1+S)]^{1/2}} [f_R (\r) + f_L (\r)]~,
\end{equation}
\begin{equation}
\psi_1 (\r) = {1 \over [2(1-S)]^{1/2}} [f_R (\r) - f_L (\r)]~, 
\end{equation}
where $f_{R,L}=\sqrt{\alpha^3 /\pi} e^{-\alpha\vert r-R_{R,L} \vert}$ 
are 1s atomic orbitals, $S$ is the overlap integral (varying from
$0$ to $1$) and
$\alpha$ is a variational parameter (varying from $1$ to $1.66$).
The two-body wave function $\phi$ can be approximated by
\begin{equation}
\phi(\r,\rp)= g_0\psi_0(\r)\psi_0(\rp) + g_1\psi_1(\r)\psi_1(\rp)~.
\end{equation}
Then
\begin{equation}
g_0=\frac{1}{\sqrt{2}} [ 1 + \frac{\kappa}
{\sqrt{1+\kappa^2}}]^{1/2},
\end{equation}
\begin{equation}
g_1=-\frac{1}{\sqrt{2}} [ 1 - \frac{\kappa}
{\sqrt{1+\kappa^2}}]^{1/2},
\end{equation}
where 
$\kappa$ is a function of $S$ and of the integrals
\begin{equation}
U=< f_Lf_L~| \frac{1}{r_{12}}|~f_Lf_L >,
\end{equation}
\begin{equation}
V=< f_Lf_R~| \frac{1}{r_{12}}|~f_Lf_R >,
\end{equation}
\begin{equation}
t=< f_Lf_L~| \frac{1}{r_{12}}|~f_Lf_R >,
\end{equation}
\begin{equation}
J=< f_Lf_L~| \frac{1}{r_{12}}|~f_Rf_R >.
\end{equation}
For large $d$, $S\approx 0$, $\kappa=2t/(U-V)$.
Therefore, when $d \rightarrow \infty$,
$g_0=-g_1=1/\sqrt{2}$ and
\begin{equation}
\phi(\r,\rp)= g_0(f_R(\r)f_L(\rp)+f_L(\r)f_R(\rp)).
\end{equation}
This means that the correlation effects drive the electrons 
back on their own atoms like in the Heitler-London ansatz.

For the linear chain molecule
H$_4$, the $|~\psi_i>$ ($i=0,1,2,3$), have $i$ nodes. 
In momentum space, $<\p~|~\psi_0>$ is peaked at $p=0$,
but the $<\p~|~\psi_i>$ (for $i=1,2,3$) are peaked at higher momenta.
When $d$ is small, only $|~\psi_0>$ and $|~\psi_1>$ are occupied, 
while, in the limit $d \rightarrow \infty $, 
the SGA method yields $g_0=g_1=-g_2=-g_3$. 
In the Hartree-Fock approximation \cite{vos}, the momentum density
$n(p)$ of the chain H$_{32}$ 
is more similar to that of a 
free-electron gas, with a given Fermi momentum $p_F$,
rather than that of the hydrogen atom. 
However, when the occupation number can vary, one expects 
$n(p)$ to develop 
high momentum tails.
Recent experiments probing the electron momentum distribution in
simple metals \cite{sakurai,alfred} have observed similar tails.

The Homogeneous Electron Gas (HEG) is another interesting
limit for solids.
In this system, the plane waves are the natural orbitals and
the total energy per particle $\varepsilon=E/N$ is a function
of the density parameter $r_s$ (i.e. the radius of the volume
taken by one electron). 
The difference between the interacting and free
HEG momentum densities for different $r_s$ 
yields the Lam-Platzman correction \cite{phil}
within the density functional theory.

Cs\'anyi and Arias \cite{gabor} computed the GU energy functional
in the HEG and minimized the result with respect to the
the occupancy $n(k)$.
At high density (small $r_s$),
the result seems to reproduce the correct RPA limit 
and $n(k)$ has a Daniel-Vosko like 
momentum dependence \cite{vosko}.
However 
when $r_s=1$, one finds
$\varepsilon=0.546$ (a.u.), while the Diffusion Quantum Monte Carlo
gives $\varepsilon=0.596$ (a.u.)
\cite{gerardo}.
This is a quite surprising result, since a variational result should be
always greater than the exact energy. 
The reason is that the 
two-particle reduced density matrix $\sigma$ has been varied over too large
class of functions: the restriction to $N$-representable $\sigma$ has not
been imposed. In other words, one cannot find a many-body state yielding
this $\sigma$.
The AGP is by definition $N$-representable.
Therefore, it provides a general variational scheme for many-electron
system.
When the AGP is applied to the interacting (with Coulomb repulsion) HEG, 
one finds 
the independent particle occupation \cite{alexandrov}. 
However, 
correlation effects in the particle occupation 
may appear in an
inhomogeneous electron gas.

\begin{figure}[htb]
\centering
\epsfxsize=\linewidth
\epsffile{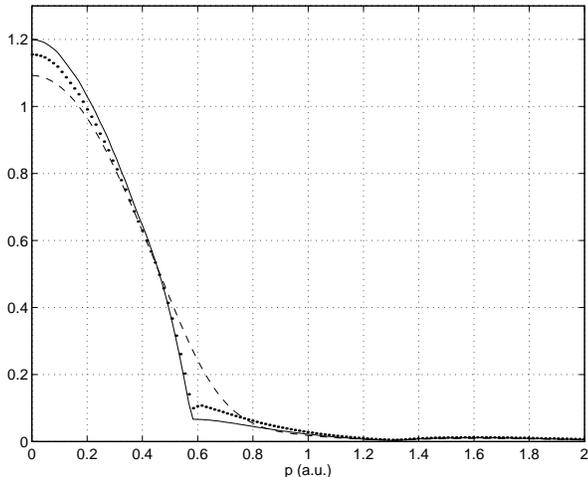}
\caption{
Total valence-electron Compton profiles of Li along (1 0 0).
The solid line is the LDA calculation, the dotted line is 
the LDA with Lam-Platzman corrections and the dashed line is the AGP
with $|\Delta(\k)|=0.1$ a.u..
}
\label{Fig}
\end{figure}

A crucial question is whether or not the fixed node approximation,
used by the Quantum Monte Carlo (QMC) simulations, gives a significant
momentum density error in realistic extended systems.
A recent QMC calculation
for solid Li \cite{claudia}
corrects about 30 \%
the discrepancy between the experimental Compton profiles
\cite{sakurai} 
and the density functional result.
The fixed node approximation might be a cause of
the remaining discrepancy.
The Lam-Platzman correction, which is in the same nodal structure,
gives a similar result  (see Fig.~\ref{Fig}).
It is therefore worthwhile to investigate schemes
beyond the free fermion nodal
structure like the AGP. In solid Li
one can approximate the natural orbitals by the Kohn-Sham orbitals  
\cite{kubo} and  
do the following BCS ansatz 
for the occupation amplitudes \cite{gros,barnes}
\begin{equation}
g(\k)=\frac{\Delta(\k)}
{\epsilon(\k)+\sqrt{\epsilon^2(\k)+\Delta^2(\k)}}.
\end{equation}
The band energy $\epsilon(\k)$ is zero at the Fermi level
and $\Delta(\k)$ 
can be either calculated 
variationally or fitted to the experiment. 
Fig.~\ref{Fig} shows that important correlation effects
can be observed in the Li Compton profile if $|\Delta(\k)|$
is about $0.1$ a.u.. 

In conclusion, the present paper presents a 
total energy functional of natural orbitals.
The method goes beyond the Slater determinant 
nodal structure. For 2 electron systems, it
is equivalent to a configuration interaction calculation. 
It can capture important 
correlation effects in the electron momentum density calculation.
The knowledge of these effects is crucial for a proper
interpretation of the experimental spectra.

I gratefully acknowledge useful discussions with Phil Platzman and
Arun Bansil. I thank Cyrus Umrigar for sending me ref. \cite{stefan}.
I also acknowledge kind hospitality and discussions
with Sudip Chakravarty and Steven Kivelson at UCLA.

\end{document}